\documentclass{PoS}

\usepackage{amsmath}

\newcommand{\ima}{{\mbox{Im}\,}}

\DeclareMathOperator*{\res}{Res}

\title{Chiral Extrapolations of light resonances from dispersion relations and Chiral Perturbation Theory}

\ShortTitle{Chiral Extrapolations of light resonances 
  from dispersion relations and Chiral Perturbation Theory}

\author{\speaker{Guillermo RÍOS MÁRQUEZ}%
         \\
        Universidad Complutense de Madrid\\
        E-mail: \email{griosmar@fis.ucm.es}}

\author{Ángel GÓMEZ NICOLA\\
        Universidad Complutense de Madrid\\
        E-mail: \email{gomez@fis.ucm.es}}

\author{Christoph HANHART\\
        Institut f\"ur Kernphysik and J\"ulich Center for Hadron
        Physics, Forschungzentrum J\"ulich GmbH\\
        E-mail: \email{c.hanhart@fz-juelich.de}}

\author{José Ramón PELÁEZ SAGREDO\\
        Universidad Complutense de Madrid\\
        E-mail: \email{jrpelaez@fis.ucm.es}}

      \abstract{We review our recent study of the pion mass dependence
        of the $\rho$ and $\sigma$ resonances, generated from one-loop
        $SU(2)$ Chiral Perturbation Theory (ChPT) with the Inverse
        Amplitude Method (IAM) which was modified to properly account
        for the Adler zero. The method is based on analyticity,
        elastic unitarity and ChPT at low energies, thus yielding the
        pion mass dependence of the resonance pole positions
        from the ChPT series up to a given order.
        We find that the
        $\rho\pi\pi$ coupling constant is almost $m_\pi$ independent
        and that our prediction compare well with some recent lattice
        results for the $\rho$ mass. These findings 
        may be relevant for
        studies of the meson spectrum and form factors on the lattice.}

\FullConference{International Workshop on 
  Effective Field Theories: from the pion to the upsilon\\
		 2-6 February 2009\\
		 Valencia, Spain}

\begin{document}

\section{Introduction}

Light hadron spectroscopy at low energies lies beyond the realm of
perturbative QCD. Although lattice QCD provides, in principle, a
rigorous way to extract non--perturbative quantities from QCD,
 current lattice results are typically still obtained for
relatively high quark masses. Thus, in order to make contact with
experiment, appropriate extrapolation formulas need to be derived. This
is typically done by using Chiral Perturbation Theory
(ChPT)\cite{chpt1}, which provides a model independent description of
the dynamics of the lightest mesons, namely, the pions, kaons and
etas, which are identified with the Goldstone Bosons (GB) associated
to the QCD spontaneous Chiral Symmetry breaking. 
Hence, ChPT is built out of
only those fields, as a low energy expansion of a Lagrangian whose
terms respect all QCD symmetries, and in particular its symmetry
breaking pattern. Actually, this chiral expansion becomes a
series in momenta and meson masses, generically $O(p^2/\Lambda^2)$,
when taking into account systematically the small quark masses of the
three lightest flavors that can be treated perturbatively.  
The chiral expansion scale is $\Lambda\equiv 4 \pi f_\pi$, where
$f_\pi$ denotes the pion decay constant.
ChPT is renormalized order by order by absorbing loop
divergences in the renormalization of parameters of higher order
counterterms, known as low energy constants (LEC), which
are the coefficients of the energy and mass expansion,
so that \emph{they have no quark mass dependence}.
Their  values depend
on the specific QCD dynamics, and have to be determined either from
experiment or from lattice QCD.

The relevant remark for us is that, thanks to the fact that ChPT has
the same symmetries as QCD and that it couples to different kind of
currents through the same pattern, the ChPT expansion
provides a {\it systematic and model independent} 
description of how the observables depend on some QCD
parameters, like the quark masses, and
this can be implemented systematically up to 
the desired order in the ChPT expansion.


We review here our recent derivation of a modified version of the IAM
\cite{modIAM} based on dispersion theory, unitarity and ChPT to
next--to--leading order (NLO), which also accounts properly for the
Adler zero.  Within this approach we are able to 
predict the quark mass dependence
of the $\sigma$ and $\rho$ mesons \cite{Hanhart:2008mx},
thus providing an explicit representation of the 
LECs appearing in the ChPT analysis
of the vector meson mass of Refs.~\cite{bruns,mainz}.

In this work we focus on the two lightest resonances
of QCD, the $\rho$ and the $\sigma$.
It is therefore enough to work with the two lightest quark
flavors $u,d$ in the isospin limit with a mass
$\hat m=(m_u+m_d)/2$. Since $m_\pi$ is given by 
$m_\pi^2\sim \hat m+...$ \cite{chpt1},
studying the $\hat m$ dependence is equivalent to study
the $m_\pi$ dependence.

\section{Unitarization and dispersion theory}

The $\sigma$ and $\rho$ resonances appear as poles on the
second Riemann sheet of the $(I,J)=(0,0)$ and $(1,1)$
$\pi\pi$ scattering partial waves of isospin $I$ 
and angular momentum $J$, respectively. 
Unitarity implies for these partial waves, 
and physical values of $s$ below inelastic thresholds, 
that
\begin{equation}
  \label{unit}
  \ima t(s)=\sigma (s) \vert t(s)\vert^2 
  \;\;\Rightarrow\;\; 
  \ima\frac1{t(s)}=-\sigma(s),\qquad {\rm with}
  \quad \sigma(s)=2p/\sqrt{s},
\end{equation}
where s is the Mandelstam variable and $p$ is the
center of mass momentum. Consequently, the imaginary part
of the inverse amplitude is known exactly. Hoewever, 
ChPT ampitudes, being an expansion
$t\simeq t_2+t_4+\cdots$, with
$t_k=O(p^k)$, can only satisfy
Eq. (\ref{unit}) perturbatively
\begin{equation}
  \label{unitpertu}
  \ima t_2(s)=0,\;\;\;
  \ima t_4(s)=\sigma(s)\vert t_2(s)\vert^2,\;\;\dots
  \quad\Rightarrow\quad
  \ima \frac{t_4(s)}{t_2^2(s)}=\sigma(s),
\end{equation}
and cannot generate poles. Therefore the resonance region lies
beyond the reach of standard ChPT. This region however, can be
reached combining ChPT with dispersion theory either for the
amplitude~\cite{gilberto}
or for the inverse amplitude through the
IAM~\cite{Truong:1988zp,Dobado:1996ps,Guerrero:1998ei}.

Here we review our recent derivation \cite{modIAM} of a modified
IAM formula to properly account for the Adler zero region.
The main ingredient is a dispersion relation for the inverse
amplitude, whose analytic structure consist on a 
right cut from $s=4m_\pi^2$ to $s=\infty$,
a left cut from $s=-\infty$ to $s=0$, and possible poles
comming from zeros of the amplitude. Indeed, for the scalar waves the
amplitude vanishes at the so called Adler zero, $s_A$, that lies
in the real axis below threshold, thus within the ChPT region of
applicability. Its position can be approximated from ChPT, i.e.,
$s_A=s_2+s_4+\cdots$, where $t_2$ vanishes at $s_2$, $t_2+t_4$
at $s_2+s_4$, etc.

We now write a once substracted dispersion relation for the
inverse amplitude, where the substraction point has been chosen
to be the Adler zero, $s_A$:
\begin{equation}
  \label{1/tdisp}
  \frac1{t(s)}=\frac{s-s_A}{\pi}
  \int_{RC}ds'\frac{\ima 1/t(s')}{(s'-s_A)(s'-s)}+
  LC(1/t)+PC(1/t),
\end{equation}
where $LC$ stands for a similar integral over the left cut and
$PC$ stands for the contribution of the pole at the Adler zero,
which reads:
$$
PC(1/t)=-(s-s_A)\res_{s'=s_A}
\left(
  \frac{1/t(s')}{(s'-s)(s'-s_A)}
\right)=
\frac1{t'(s_A)(s-s_A)}-\frac{t''(s_A)}{2t'(s_A)^2}.
$$
The different terms in 
Eq. (\ref{1/tdisp}), can be evaluated in the following
way:
\begin{itemize}
\item The right cut can be \emph{exactly} evaluated
  taking into account the elastic unitarity condition
  Eqs.~(\ref{unit}),~(\ref{unitpertu}),
  $\ima 1/t(s')=-\sigma(s')=-\ima t_4(s')/t_2^2(s')$,
  for $s'\in (4m_\pi^2,\infty)$.

\item The pole contribution only involves amplitude
  derivatives evaluated at the Adler zero,
  \emph{which is a low energy point}, so they can be
  can be safely approximated with ChPT.

\item The left cut, which is $1/(s'-s)$ suppressed for $s$
  values near the physical region, is 
  \emph{weighted at low energies},
  so it is appropiate to approximate it with ChPT.

\item Finally, in the cut contributions we approximate
  $(s-s_A)/(s'-s_A)\simeq (s-s_2)/(s'-s_2)$, which is
its LO chiral expansion and a remarkably good approximation
as long as $s'$ is  is sufficiently far from $s_2$ and $s_A$,
which is indeed the case for the cut integrals.
\end{itemize}
Altogether we are able to write Eq. (\ref{1/tdisp})
as
\begin{equation}
  \label{dispeval}
  \frac1{t(s)}\simeq 
  \underbrace{-RC(t_4/t_2^2)-LC(t_4/t_2^2)}
  _{-t_4(s)/t_2^2(s)+PC(t_4/t_2^2)}
  +PC(1/t)=
  -\frac{t_4(s)}{t_2^2(s)}+PC(t_4/t_2^2)+PC(1/t),
\end{equation}
where we have taken into account that a once subtracted 
dispersion relation can be written for $t_4/t_2^2$,
the substraction point being $s_2$:
\begin{equation}
  \label{t4/t2disp}
  \frac{t_4(s)}{t_2^2(s)}=\frac{s-s_2}{\pi}
  \int_{RC}ds'\frac{\ima\, t_4(s')/t_2^2(s')}{(s'-s_2)(s'-s)}
  +LC(t_4/t_2^2)+PC(t_4/t_2^2),
\end{equation}
where $PC(t_4/t_2^2)$ stands for the contribution of the
triple pole at $s_2$.
From Eq. (\ref{dispeval}) one easily arrives at the
modified IAM (mIAM) formula \cite{modIAM}:
\begin{equation}
  \label{mIAM}
  t^{mIAM}(s)\simeq \frac{t_2^2(s)}{t_2(s)-t_4(s)+A(s)},
  \quad A(s)=t_4(s_2)-\frac{(s_2-s_A)(s-s_2)}{s-s_A}
  \Big(t'_2(s_2)-t'_4(s_2)\Big).
\end{equation}
The standard IAM formula is recovered for $A(s)=0$, which holds
exactly for all partial waves except the scalar ones. In the original
IAM derivation \cite{Truong:1988zp,Dobado:1996ps} $A(s)$ was
neglected, since it formally yields a NNLO contribution and is
numerically very small, except near the Adler zero, where it diverges.
However, if $A(s)$ is neglected, the IAM Adler zero occurs at $s_2$,
correct only to LO, it is a double zero instead of a simple one, and
a spurious pole of the amplitude appears close to the Adler zero on
both the first and second sheet. All of
these caveats are removed with the mIAM, Eq. \eqref{mIAM}. The
differences in the physical and resonance region between the IAM and
the mIAM are less than 1\%. However, as we will see, for large $m_\pi$
the $\sigma$ poles appear as two virtual poles
on the second sheet below threshold, one of
them moving towards zero and approaching the Adler zero region, where
the IAM fails. Thus, we will use for our calculations the mIAM,
although it is only relevant for the mentioned second $\sigma$ pole,
and only when it is very close to the spurious pole.

In summary, there are no model dependences in the approach, 
but just approximations to a given order in ChPT.
Let us remark that, in our derivation, ChPT has been used
always \emph{at low energies} to evaluate parts of a
dispersion relation, whose elastic unitarity cut is
taken into account exactly. Thus, the IAM formula is
reliable up to energies where inelasticities become
important, even though ChPT does not converge at those
energies, because ChPT is not being used there.
This argument also holds when the $\sigma$ moves
near threshold, because the Adler zero, which is the
point where ChPT is used, is still at low energies
far away from the pole. Note that the virtual pole that goes down
in energies is on the second sheet, so does not prevent the
use ChPT at the Adler zero, on the first sheet.
Remarkably, the simple formula of the elastic mIAM, 
Eq.(\ref{mIAM}), (or the standard IAM one),
while reproducing the ChPT expansion at low energies,
is also able to generate both the  $\rho$ and $\sigma$ 
resonances with values of the LECs compatible with 
standard ChPT \cite{Guerrero:1998ei}.
In other words, the IAM generates the 
poles \cite{Dobado:1996ps}
associated to these resonances in the second Riemann
sheet. The fact that resonances are 
\emph{not introduced by hand} but generated
from first principles and data, is relevant because the 
existence and nature of scalar resonances is the
subject of a long-lasting intense debate.

To be precise, the IAM, when reexpanded, reproduces the
ChPT series up to the order to which the input amplitude was evaluated
and, in particular, the quark mass dependence agrees with that of ChPT 
up to that order.
A few of the higher order terms are produced correctly
by the unitarization but not the complete series--- for
a discussion of this issue for the scalar pion form factor see 
Ref.\cite{Gasser:1990bv}. However, 
the formalism just described still provides
us with a fair estimate of the quark mass dependence of the resonance
properties. 

\section{Results}
By changing $m_\pi$ in the amplitudes we see how the poles
generated with the IAM evolve. 
We will use the LECs values 
$10^3l^r_3=0.8\pm 3.8$ and $10^3l^r_4=6.2\pm 5.7$ 
from \cite{chpt1} and fit the mIAM to data 
up to the resonance region to find
$10^3l^r_1=-3.7\pm 0.2$ and $10^3l^r_2=5.0\pm 0.4$. These LECs 
are evaluated at $\mu=770$ MeV.

The values of $m_\pi$ considered should fall within the ChPT
range of applicability and allow for some elastic $\pi\pi$
regime below $K\bar K$ threshold. Both criteria are
satisfied if $m_\pi\leq 500$ MeV, since $SU(3)$ ChPT
still works with such kaon masses, and because for
$m_\pi\simeq 500$ MeV, the kaon mass becomes $\simeq 600$, leaving
200 MeV of elastic region. Of course, we expect higher order
corrections, which are not considered here, to become more
relevant as $m_\pi$ is increased. Thus, our results become less
reliable as $m_\pi$ increases due to the $O(p^6)$ 
corrections which we have neglected

\begin{figure}[t]
  \begin{center}
    \hbox{
      \includegraphics[scale=0.25,angle=0]{polesNew.eps}
      \includegraphics[scale=0.60,angle=0]{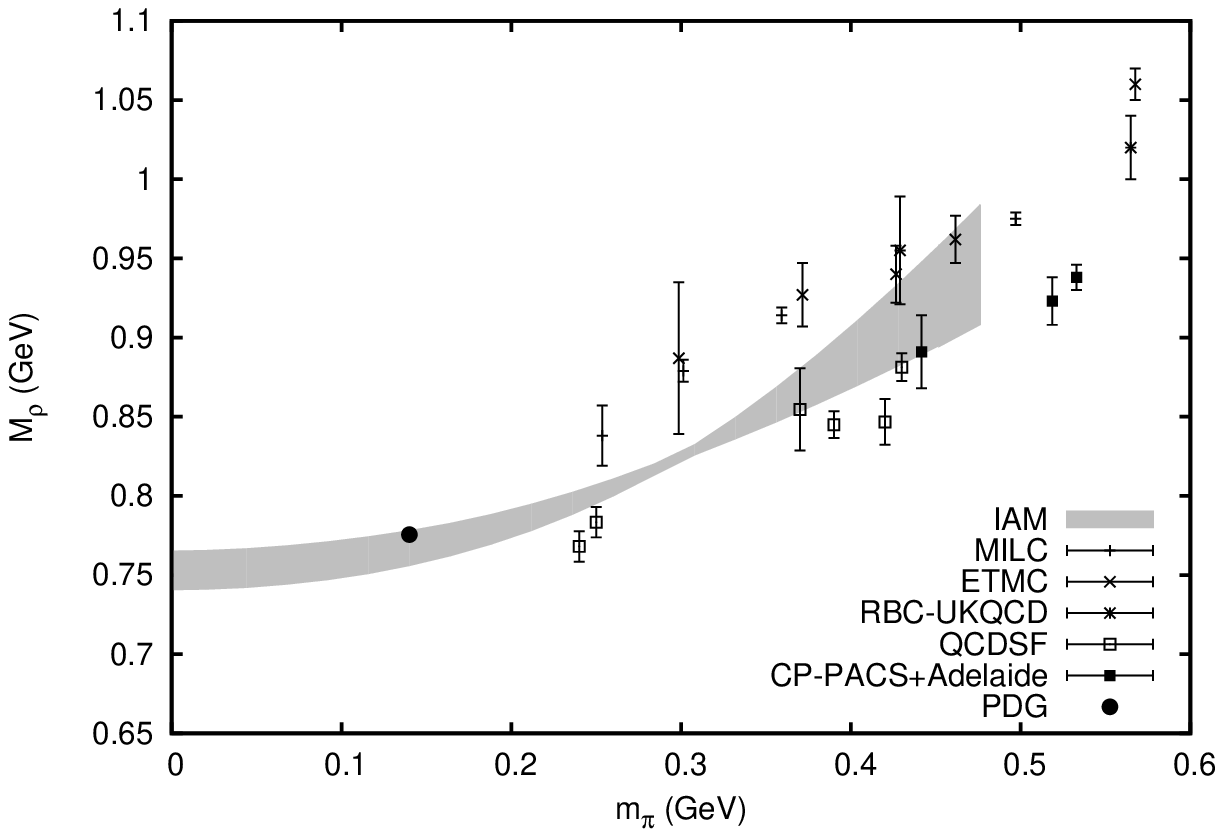}
    }
    \caption{ {\bf Left:} Movement of the $\sigma$ (dashed lines) 
      and $\rho$ (dotted
      lines) poles for increasing $m_\pi$ (direction indicated by the
      arrows) on the second sheet.  The filled (open) boxes denote the
      pole positions for the $\sigma$ ($\rho$) at pion masses $m_\pi=1,\
      2,$ and $3 \times m_\pi^{\rm phys}$, respectively. For
      $m_\pi=3m_\pi^{\rm phys}$ three poles accumulate in the plot 
      very near the $\pi\pi$ threshold. Note that all poles are
      always far enough from the Adler zero (circles).
      {\bf Right:} Comparison of our results for the $M_\rho$
      dependence on $m_\pi$ with some recent lattice results
      from \cite{lattice1}. The grey band covers only the
      error coming from the LECs uncertainities.}
      \label{poles}
  \end{center}
\end{figure}

Fig. \ref{poles} (left) shows the evolution of the $\sigma$ and $\rho$
pole positions as $m_\pi$ is increased. In order to see the pole
movements relative to the two pion threshold, which is also increasing,
all quantities are given in units of $m_\pi$, so the threshold is
fixed at $\sqrt{s}=2$. Both poles moves closer to threshold and
they approach the real axis. The $\rho$ poles reach the real axis
as the same time that they cross threshold.
One of them jumps into the first sheet and stays below
threshold in the real axis as a bound state, while its conjugate
partner remains on the second sheet practically at the very same
position as the one in the first. In contrast, the $\sigma$
poles go below threshold with a finite imaginary part before they
meet in the real axis, still on the second sheet, becoming
virtual states. As $m_\pi$ is increased further, one of the poles
moves toward threshold and jumps through the branch point to the
first sheet and stays in the real axis below threshold, very
close to it as $m_\pi$ keeps growing. The other $\sigma$ pole moves
down in energies further from threshold and remains 
on the second sheet.
This analytic structure, with
two very asymmetric poles in different sheets for a scalar wave, could be
a signal of a prominent molecular component \cite{Weinberg,baru},
at least for large pion masses.
Similar pole movements have been also found within quark models
\cite{vanBeveren:2002gy} and in finite density analysis 
\cite{FernandezFraile:2007fv}.

\begin{figure}[t]
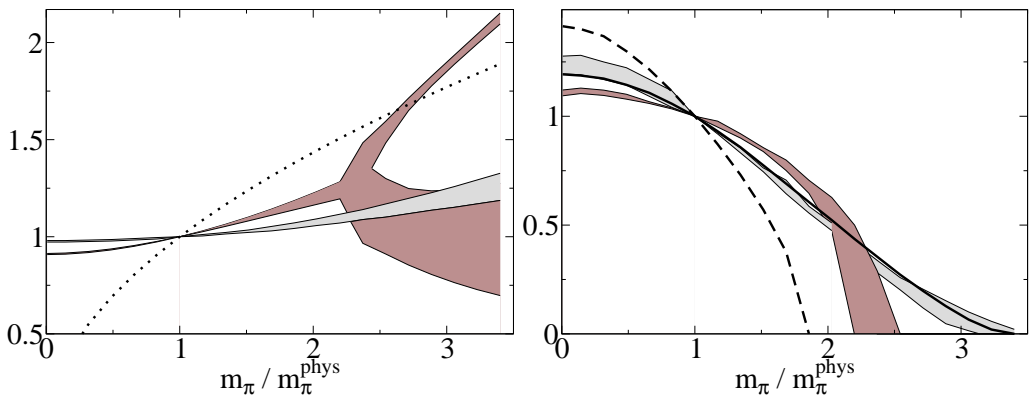
 
  \begin{center}
    \hbox{
      \includegraphics[scale=0.25,angle=0]{MryMsNew.eps}
      \includegraphics[scale=0.25,angle=0]{WryWsNew.eps}
    }
    \caption{\label{massandwidth} $m_\pi$ dependence of resonance masses (left)
      and widths (right) in units of the physical values. 
      In both panels the dark (light) band shows the results for
      the $\sigma$ ($\rho$). The width of the bands reflects 
      the uncertainties
      induced only from the uncertainties in the LECs. The dotted line 
      shows the $\sigma$ mass dependence
      estimated in Ref.~\cite{Jeltema:1999na}. The dashed (continuous)
      line shows the $m_\pi$ dependence of the $\sigma$ ($\rho$) 
      width from the change of phase space only, assuming
      a constant coupling of the resonance to $\pi\pi$.}
  \end{center}
\end{figure}

In Fig. \ref{massandwidth} (left) we show the $m_\pi$ dependence
of $M_\rho$ and $M_\sigma$ (defined from the pole position
$\sqrt{s_{pole}}=M-i\Gamma /2$), normalized to their physical values.
The bands cover the LECs uncertainties. We see that both masses
grow with $m_\pi$, but $M_\sigma$ grows faster than $M_\rho$. 
Below $m_\pi\simeq 330$  MeV we only show one line because the two
conjugate $\sigma$ poles have the same mass. Above 330 MeV, these
two poles lie on the real axis with two different masses. The
heavier pole goes towards threshold and around $m_\pi\simeq 465$ MeV
moves into the
first sheet. Note also that the
$m_\pi$ dependence of $M_\sigma$ is much softer than suggested in
\cite{Jeltema:1999na}, shown as the dotted line, which in addition
does not show the two virtual poles.

In the right panel of Fig. \ref{massandwidth} we show the $m_\pi$
dependence of $\Gamma_\rho$ and $\Gamma_\sigma$ normalized to their
physical values, where we see that both widths become smaller. 
We compare this decrease with the expected reduction from
phase space as the resonances approach the $\pi\pi$ threshold.
We find that $\Gamma_\rho$ follows very well
this expected behavior, which
implies that the $\rho\pi\pi$ coupling is almost $m_\pi$ independent.
In contrast $\Gamma_\sigma$ shows a different behavior from the
phase space reduction expectation. This suggest a strong $m_\pi$
dependence of the $\sigma$ coupling to two pions, necessarily
present for molecular states \cite{baru,mol}.

Fig. \ref{poles} (right) shows our results for the $\rho$ mass
(here defined as the point where the phase shift crosses $\pi/2$,
except for those $m_\pi$ values where the $\rho$ becomes a bound
state, where it is defined then from the pole position)
dependence on $m_\pi$ compared with some recent lattice 
results~\cite{lattice1}, where we also quote the
PDG value for the $\rho$ mass. 
Taking into
account the incompatibilities within errors between
different lattice colaborations, we find a qualitaive good
agreement with the lattice resuts. Also,
we have to take into account that the $m_\pi$ dependence
in our approach is correct only up to NLO in ChPT, and
we expect higher order corrections to be important for
large pion masses.

\section{Summary}

We have reviewed our derivation of a modified version of the IAM
\cite{modIAM}, which is derived from the first principles of
analyticity, unitarity, and ChPT at low energies and has the correct
behaviour around the Adler zero. It is able to generate the $\sigma$
and $\rho$ resonance poles without any a priori assumptions, and
yields the correct dependence on the pion mass up to NLO in ChPT. We
have predicted the evolution of the resonance pole positions with
increasing pion mass \cite{Hanhart:2008mx} and have seen that both
resonances become bound states. We have also shown that the
$\rho\pi\pi$ coupling constant is almost $m_\pi$ independent and we
have found a qualitative agreement with some lattice results for the
$\rho$ mass evolution with $m_\pi$.  These findings might be
relevant for studies of the meson spectrum and form factors---see
Ref.~\cite{FFpaper}---on the lattice.


\begin{thebibliography}{99}
  \bibitem{chpt1}
    S. Weinberg, Physica {\bf A96} (1979) 327.
    J.~Gasser and H.~Leutwyler,
    Annals Phys.\  {\bf 158} (1984) 142;
Nucl.\ Phys.\ B {\bf 250} (1985) 465.


\bibitem{modIAM} A. G\'omez Nicola, J.R. Pel\'aez and G. R\'{\i}os,
Phys. Rev. D {\bf 77} (2008) 056006.

\bibitem{Hanhart:2008mx}
  C.~Hanhart, J.~R.~Pelaez and G.~Rios,
  Phys.\ Rev.\ Lett.\  {\bf 100}, 152001 (2008)

\bibitem{bruns}
  P.~C.~Bruns and U.-G.~Mei\ss ner,
  Eur.\ Phys.\ J.\  C {\bf 40} (2005) 97

\bibitem{mainz}
D.~Djukanovic, J.~Gegelia, A.~Keller and S.~Scherer,
  [arXiv:hep-ph/0902.4347].


\bibitem{gilberto}
 I.~Caprini et al.,
  Phys.\ Rev.\ Lett.\  {\bf 96} (2006) 132001


\bibitem{Truong:1988zp}
T.~N.~Truong,
Phys.\ Rev.\ Lett.\  {\bf 61} (1988) 2526.
Phys.\ Rev.\ Lett.\ {\bf 67}, (1991) 2260;
A. Dobado et al., Phys.\ Lett.\ {\bf B235} (1990) 134.




\bibitem{Dobado:1996ps}
A.~Dobado and J.~R.~Pel\'aez,
Phys.\ Rev.\ D {\bf 47} (1993) 4883;
Phys.\ Rev.\ D {\bf 56} (1997) 3057.
\bibitem{Guerrero:1998ei}
F.~Guerrero and J.~A.~Oller,
Nucl.\ Phys.\ B {\bf 537} (1999) 459
[Erratum-ibid.\ B {\bf 602} (2001) 641].
  J.~R.~Pel\'aez,
  Mod.\ Phys.\ Lett.\ A {\bf 19}, 2879 (2004)
A.~G\'omez Nicola and J.~R.~Pel\'aez,
Phys.\ Rev.\ D {\bf 65} (2002) 054009 and
AIP Conf.\ Proc.\  {\bf 660} (2003) 102.
[hep-ph/0301049].

 \bibitem{Gasser:1990bv}
  J.~Gasser and U.-G.~Mei\ss ner,
  Nucl.\ Phys.\  B {\bf 357} (1991) 90.

 \bibitem{lattice1}
  Ph.~Boucaud {\it et al.}  [ETM Collaboration],
  Phys.\ Lett.\  B {\bf 650}, 304 (2007)
  C.~Allton {\it et al.}  [RBC and UKQCD Collaborations],
  Phys.\ Rev.\  D {\bf 76}, 014504 (2007)
  C.~W.~Bernard {\it et al.},
  Phys.\ Rev.\  D {\bf 64}, 054506 (2001)
  C.~R.~Allton {\it et al.}
  Phys.\ Lett.\  B {\bf 628}, 125 (2005)
  M.~Gockeler {\it et al.}
  [QCDSF Collaboration],

 \bibitem{Weinberg}
 D.~Morgan,
  Nucl.\ Phys.\  A {\bf 543} (1992) 632;   D.~Morgan and M.~R.~Pennington,
  Phys.\ Rev.\  D {\bf 48} (1993) 1185.


 \bibitem{baru}
  V.~Baru et al.,
   Phys.\ Lett.\  B {\bf 586} (2004) 53.

\bibitem{vanBeveren:2002gy}
  E.~van Beveren et al.,
  AIP Conf.\ Proc.\  {\bf 660}, 353 (2003);
  Phys.\ Rev.\  D {\bf 74}, 037501 (2006).

\bibitem{FernandezFraile:2007fv}
  D.~Fernandez-Fraile, A.~Gomez Nicola and E.~T.~Herruzo,
  Phys.\ Rev.\  D {\bf 76}, 085020 (2007)

\bibitem{Jeltema:1999na}
  T.~E.~Jeltema and M.~Sher,
  Phys.\ Rev.\  D {\bf 61} (2000) 017301

\bibitem{mol}
S. Weinberg, Phys. Rev. {\bf
130}, 776 (1963);   Y.~Kalashnikova et al.,
  Eur.\ Phys.\ J.\  A {\bf 24} (2005) 437.

\bibitem{FFpaper}
  F.~K.~Guo, C.~Hanhart, F.~J.~Llanes-Estrada and U.~G.~Meissner,
  [arXiv:hep-ph/0812.3270].
 
\end{thebibliography}
\end{document}